\documentclass[
reprint,
amsmath,amssymb,
 aps,
noeprint
]{revtex4-2}

\usepackage{graphicx}
\usepackage{dcolumn}
\usepackage{bm}

\newcommand{\Rb}{$^{87}$Rb}
\newcommand{\ket}[1]{\left|#1\right\rangle}
\newcommand{\bra}[1]{\left\langle#1\right|}
\newcommand{\ketbra}[2]{\left|#1\right\rangle\left\langle#2\right|}
\newcommand{\wl}{\omega_{\rm L}}
\newcommand{\mF}{m_{\rm F}}

\newcommand{\mB}{\mu_{\rm B}}

\newcommand{\beginsupplement}{%
        \setcounter{table}{0}
        \renewcommand{\thetable}{S\arabic{table}}%
        \setcounter{figure}{0}
        \renewcommand{\thefigure}{S\arabic{figure}}%
        \renewcommand{\theequation}{S\arabic{figure}}
     }

\begin{document}

\preprint{APS/123-QED}

\title{Engineering atomic polarization with microwave-assisted optical pumping}%

\author{A.\ Tretiakov}
\affiliation{Dept.\ of Physics, University of Alberta, Edmonton AB, Canada T6G 2E1}
\author{C.\ A.\ Potts}
\affiliation{Dept.\ of Physics, University of Alberta, Edmonton AB, Canada T6G 2E1}
\author{Y.\ Y.\ Lu}
\affiliation{Dept.\ of Physics, University of Alberta, Edmonton AB, Canada T6G 2E1}

\author{J.\ P.\ Davis}
\affiliation{Dept.\ of Physics, University of Alberta, Edmonton AB, Canada T6G 2E1}
\author{L.\ J.\ LeBlanc}
\affiliation{Dept.\ of Physics, University of Alberta, Edmonton AB, Canada T6G 2E1}

\begin{abstract}
Polarized atomic ensembles play a crucial role in precision measurements.
We demonstrate a novel method of creating atomic polarization in an alkali vapor in a continuous-wave regime. The method relies on a combination of optical pumping by a laser beam and microwave transitions due to a cavity-enhanced magnetic field. With this approach, atomic internal angular momentum can be oriented along a static magnetic field at an arbitrary angle with respect to the laser beam. Furthermore, the atomic polarization depends on the microwave parameters, which can be used for microwave-to-optical transduction and microwave-controlled nonlinear magneto-optical rotation.
\end{abstract}

\maketitle


Throughout atomic physics, the light-matter interaction is used  to control  quantum states of matter, with results that range from revealing fundamental physics~\cite{Georgescu2014} to enabling precision timekeeping~\cite{Ludlow2015}.  The advent of the laser meant that the electric-dipole transitions (with their stronger couplings) could be readily accessed and precisely controlled within the atom.  In parallel,  microwave transitions, enabled by the weaker magnetic-dipole interactions such as in the ``clock'' transitions between hyperfine ground states of  alkali metal vapours, can be independently and simultaneously manipulate atomic states, providing a path towards a microwave-to-optical interface using atomic media~\cite{Lauk2020a}.

As a platform for quantum technologies, alkali-metal atomic ensembles, which naturally support microwave and optical transitions, provide a good opportunity for microwave-to-optical quantum transduction~\cite{Vewinger2007,Gard2017,Han2018,Vogt2019,Liang2019,Liu2021} and for storing and retrieving quantum information on demand~\cite{Zhang2009,Katz2018,Saglamyurek2021}.  The nonlinear effects of microwave interactions on optical properties of alkali vapors also include the transduction of classical signals~\cite{Cox2018,Deb2018,Anderson2018,Meyer2018a,Song2019,Jiao2019,Gordon2019,Simons2019,Holloway2019,Tretiakov2020,Li2021}, compact atomic clocks~\cite{Petremand2012,Pellaton2012,Bandi2014}, microwave electrometry~\cite{Sedlacek2013}, and static and microwave magnetometry~\cite{Kinoshita2013,Sun2017,Shi2018,Liu2018,Tretiakov2019}. 

In many cases, the microwave's effect on the optical properties in atomic vapors is due to \emph{atomic polarization}: a particular distribution of the atoms in the ensemble among the available quantum states, which generally have well-defined angular momenta or ``spin,'' allowing for the definition of a direction along which the polarization points and the equivalent nomenclature of a \emph{spin polarization}. Polarized atomic ensembles like these are  essential components of optically pumped atomic magnetometers~\cite{Kominis2003,Ben-Kish2010,Patton2014,Limes2020,Zhang2020}; indeed, some approaches require atomic polarization to be perpendicular to the probe laser beam~\cite{Kominis2003,Patton2014}, which is non-trivial task for a single-laser setup~\cite{Katz2019}.


\begin{figure}
\includegraphics{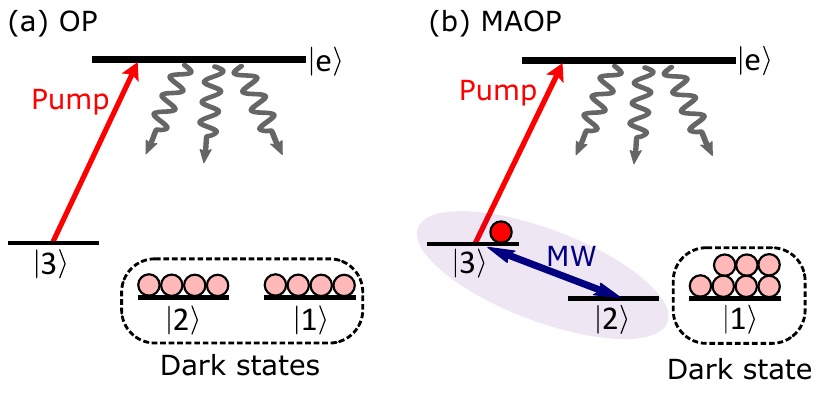}%
\caption{\label{fig:1}Principle of the optical pumping (a) and  microwave-assisted  optical pumping (b) in a four-level system.}
\end{figure}

In this Letter, we introduce a novel technique for engineering spin polarization in atomic ensembles based on microwave-assisted optical pumping (MAOP).  MAOP can be understood as an extension of microwave-optical double-resonance and similar to  optically detected magnetic resonance (ODMR) methods in solid-state systems~\cite{Wineland1980,Demtroder1996},  except here, at least one of the two ground-state levels requires several sublevels [Fig.~\ref{fig:1}]. MAOP leads to an uneven distribution of atomic populations, resulting in  atomic polarization~. We observe this polarization as a microwave-induced optical birefringence and dichroism. The dichroism can be  used to implement a polarization-selective microwave-to-optical interface, while the birefrengince leads to microwave-controlled nonlinear magneto-optical rotation~\cite{Zigdon2010} that can be applied for signal transduction based on pulse carving~\cite{Fenwick2020}.

The atomic polarization emerging from MAOP results from the particular configuration of  dark states that emerge in the presence of combined optical and microwave fields. To illustrate this principle, consider a four-level toy model shown in Fig.~\ref{fig:1}. Here, an optical field (``pump'') couples ground level $\ket{3}$ to  excited level $\ket{e}$ [Fig.~\ref{fig:1}(a)]. Ground states $\ket{1}$ and $\ket{2}$ are separated from state $\ket{3}$ by a microwave magnetic-dipole transition. We assume that the frequency separation is sufficient to make optical excitation of  $\ket{1}$ and $\ket{2}$ by the pump field negligible, rendering both states ``dark.'' Additionally assuming that there is no relaxation between the ground state when this system interacts only with the optical field,  optical pumping transfers the atomic population into the dark states $\ket{1}$ and $\ket{2}$. When an additional microwave field couples states $\ket{2}$ and $\ket{3}$,  $\ket{1}$ remains dark and  the atomic population accumulates here~[Fig.~\ref{fig:1}(b)], thus polarizing the ensemble.

\begin{figure}
\includegraphics{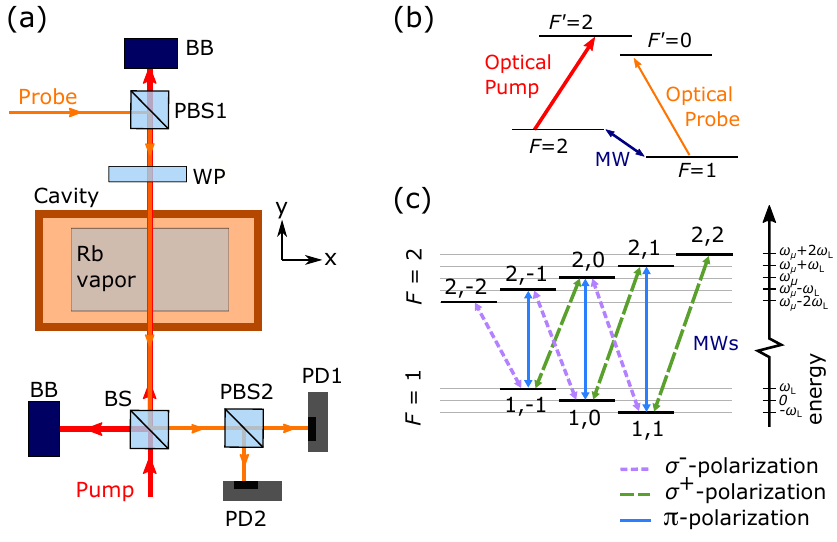}%
\caption{\label{fig:2}(a) Schematics of the experimental setup. Microwave magnetic field vector is parallel to $x$ (b) Transitions used in the experiment. Optical transitions correspond to the D2~line.  (c) Possible microwave transitions between Zeeman sublevels labelled $\ket{F,m_{\rm F}}$ of the $^{87}$Rb ground state. The transitions whose lines cross (in an ``X'') have the same frequency. A rich assortment of atomic polarizations that can be achieved by MAOP in this level manifold~\cite{SM}.}
\end{figure}

In our experiment, we observe MAOP in thermal $^{87}$Rb vapor, which is contained by a vapor cell enclosed in a microwave cavity~[Fig.~\ref{fig:2}(a)].  We use the TE$_{011}$ cavity mode tuned to the ground-state hyperfine transition $\ket{F=1,m_{\rm F}=0\rightarrow F=2,m_{\rm F}=0}$ at $6.834~682~610$~GHz. The cavity quality factor for this mode is $Q\approx 27,000$. The microwave field is provided by a microwave source,  with a power of 100~$\mu$W at the source output, without additional amplification. The cavity has a pair of holes providing optical access for the pump and probe laser beams. The pump and probe beams are provided by two different lasers. For the optical pumping we use $\ket{F=2}\rightarrow\ket{F'=2}$ transition [Fig.~\ref{fig:2}(b)] within the D2~line at around $780.2$~nm with a typical laser power of $1-2$~mW. 

For the probe, we use a typical power between $10-100\ \mu$W, and we set its frequency according to the particular measurement: in the absorptive regime, it is resonant with $\ket{F=1}\rightarrow\ket{F'=0}$ transition of the D2~line, and in the dispersive case, the probe is red-detuned from this transition.
The counterpropagating probe and pump beams are overlapped inside the cell (Fig.~\ref{fig:2}a).
We use a wave plate (WP) (either a quarter-wave plate or a half-wave plate) to control the probe polarization as it enters the cell. 
To analyze the MAOP effect on the probe polarization, a  polarizing beamsplitter (PBS2) splits the probe between two photodetectors (PD1 and PD2). 
The static magnetic field inside the cell is controlled by three pairs of Helmholtz coils along mutually perpendicular axes.

The $^{87}$Rb ground state has nine possible microwave transitions [Fig.~\ref{fig:2}(c)] with three different polarizations: $\pi$, $\sigma^{+}$, and $\sigma^{-}$.
A static magnetic field $\mathbf{B}$ inside the cell  affects both the frequency difference between the microwave transitions and the  strength of transitions corresponding to a particular polarization of the microwave field. 
The magnitudes of the static magnetic fields in our experiment are within the regime of the linear Zeeman effect, giving the frequency  separation between two neighbouring transitions by the Larmor frequency $ \omega_{\rm L} = \mu_{\rm B}|\mathbf{B}|/2\hbar$, where $\mu_{\rm B}$ is the Bohr magneton and $\hbar$ is the reduced Planck constant. Some of the transitions are shifted by the same frequency, giving two pairs of degenerate transitions. The typical values of the Larmor frequency in our experiments are on the order of $100$~kHz, which is much less than the width of the D2~line (on the order of MHz), meaning that both probe and pump couple to all Zeeman sublevels of the corresponding hyperfine levels simultaneously. 
We define the quantization axis as parallel to $\mathbf{B}$, so the polarization of the microwave field with respect to the quantization axis is determined by the angle between the microwave magnetic field vector $\mathbf{B}_{\mu}$ and $\mathbf{B}$. Here, the $\pi$ transitions correspond to the microwave vector component parallel to $\mathbf{B}$, while $\sigma^{\pm}$ transitions correspond to the microwave vector orthogonal to  $\mathbf{B}$. In our setup, the microwave field vector is parallel to the cavity axis, which is fixed, so we control the microwave polarization with respect to the quantization axis by rotating  $\mathbf{B}$ with the Helmholtz coils.

\begin{figure*}
\includegraphics{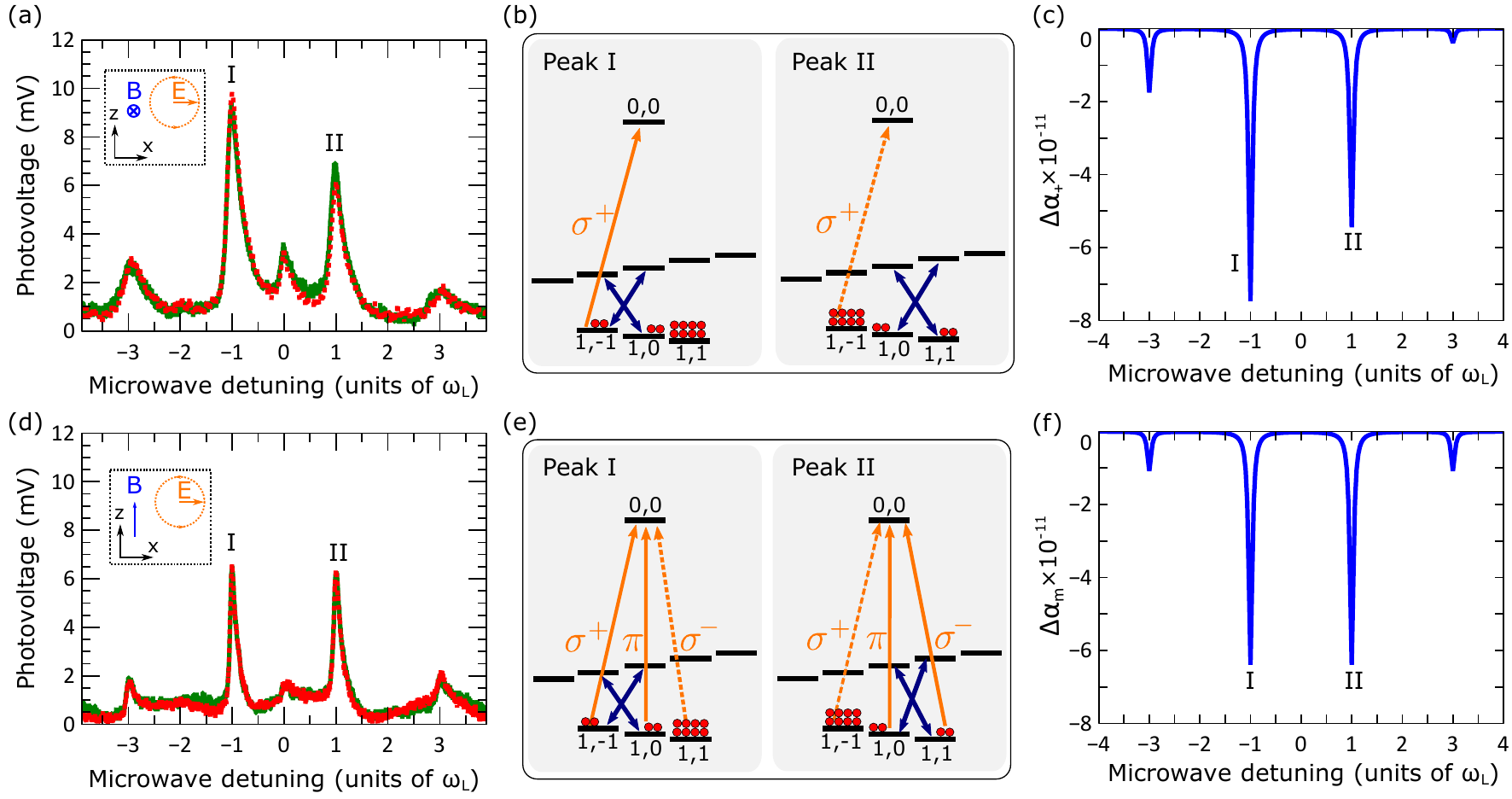}%
\caption{\label{fig:4} Absorptive measurement of MAOP. (a) Optical signal during a microwave frequency scan detected by PD1 (red dotted) and PD2 (green solid) for circularly polarized probe in case when the static magnetic field $\mathbf{B}$ is parallel to the probe beam. Inset shows the orientation of the static magnetic-field vector and the polarization of the probe electric-field vector $\mathbf{E}$ with respect to $x$~and~$y$~axes from Fig.~\ref{fig:2}(a). The probe polarization lies in the $xz$~plane. The microwave detuning is given with respect to the clock transition. 
(b) Schematics of the microwave (blue arrows) and probe (yellow arrow) transitions, and the expected relative distribution of the atomic population between Zeeman substates corresponding to peaks I and II from (a). The probe polarization is indicated with respect to the static magnetic field. The pump field is not shown. 
(c) Absorption parameter~\cite{SM} calculated for $\hbar=1$, $\Gamma_{\rm OP}=2\pi\times10^{-3}$, $\omega_{\rm L} = 2\Gamma_{\rm OP}\times10^{2}$, $\Omega_0=\Gamma_{\rm OP}\times10^{-2}$, $\theta=\pi/2$, $g_{\rm C}=0.1\omega_{\rm L}$ , and $\Gamma_{\rm th}=\Gamma_{\rm OP}$.
(d-f) are same as (a-c), except the static magnetic field is perpendicular to $xy$-plane. Microwave power at the source output is 100~$\mu$W. The frequency sweeping rate is $35.5$~Hz. The Larmor frequency $\omega_{\rm L}$ was estimated from the peak separation as $160$~kHz in (a), and $259$~kHz in (d).} 
\end{figure*}

\textbf{Absorption.}
We begin with MAOP measurements in absorptive regime [Fig.~\ref{fig:4}]. For these measurements, the probe is circularly polarized with respect to its wavevector, in which case the probe power splits equally between the two photodetectors. Fig.~\ref{fig:4}(a) shows the AC part of the photovoltages from both detectors during a linear microwave frequency scan when $\mathbf{B} = B\hat{y}$ is parallel to the probe wavevector. In this case, the probe is $\sigma^{+}$-polarized with respect to the quantization axis, while the microwave field, whose vector is orthogonal to $\mathbf{B} = B\hat{y}$, is an equal superposition of $\sigma^{+}$ and $\sigma^{-}$ polarizations. As we scan the microwave frequency, we see an increase in the probe transmission whenever the microwave is on resonance with a transition allowed by the microwave polarization. The asymmetry in the signal in Fig.~\ref{fig:4}(a)  results from the fact the microwave field  creates different atomic polarizations at different frequencies. Fig.~\ref{fig:4}(b) illustrates the population distribution for peaks I and II obtained with corresponding to opposite values of the microwave detuning with respect to the clock transition $\ket{1,0}\rightarrow\ket{2,0}$. In the first case, the microwave field simultaneously drives $\ket{1,-1}\rightarrow\ket{2,0}$ and $\ket{1,0}\rightarrow\ket{2,-1}$, in which case the MAOP transfers the atomic population to $\ket{1,1}$. Because only the $\ket{1,-1}\rightarrow\ket{0,0}$ transition is allowed for a $\sigma^+$ probe, its transmission is increased due to the reduced population of $\ket{1,-1}$. In the case of peak II, the MAOP moves the atomic population to $\ket{1,-1}$, where the probe can be absorbed, making the observed transmission  smaller than in the first case. If we reverse the bias field direction, the asymmetry of the observed signal is reversed (not shown) since in this case probe is $\sigma^{-}$ polarized and it is absorbed more strongly when  the atomic population is transferred to $\ket{1,1}$. 

Note  that  the simplified  concept of MAOP introduced in Fig.~\ref{fig:1} does not fully explain the signals observed in Fig.~\ref{fig:4}(a). For example, instead  of  peak  II,  we  would  expect  to  see  a  dip  in the  optical  transmission  since  the atomic  polarization,  in this case, provides a higher optical density compared to an unpolarized  atomic  ensemble. To resolve this, we introduced a comprehensive theoretical model describing MAOP in an ensemble with thermal relaxation~\cite{SM}. It also includes the contribution to the absorption from the velocity classes resonant with $\ket{F=1},\rightarrow\ket{F'=1}$ and $\ket{F=1}\rightarrow\ket{F'=2}$ transitions due to the Doppler effect.  With this model, we calculate the steady-state optical absorption in terms of parameter $\alpha_{\rm p}$, where $p$ denotes the particular probe polarization. Parameter $\alpha_{\rm p}$ is proportional to the absorption, and it captures the qualitative dependence on the microwave field characteristics.
Fig.~\ref{fig:4}(c) shows the absorption coefficient calculated for an ensemble dominated by thermal relaxation between the ground-state levels, and it is in a good agreement with the observed signal.

To confirm that the direction of the atomic polarization is defined by the bias magnetic field vector, we perform the same measurement with the bias field applied out-of-page ($\mathbf{B} = B\hat{z}$) [Fig.~\ref{fig:4}(d-e)], in which case, it is perpendicular to both the microwave magnetic field  and the probe wave vector. While the microwave field is still $\sigma^{\pm}$ polarized, the probe polarization now has components parallel and orthogonal to the quantization field and  couples to all three sublevels of the lower-energy ground state. Because the probe power is divided equally between $\sigma^{+}$ and $\sigma^{-}$ polarization components, the atomic polarizations corresponding to opposite microwave detunings have the same effect on optical absorption, and so the observed signal is symmetric.

\begin{figure}
\includegraphics{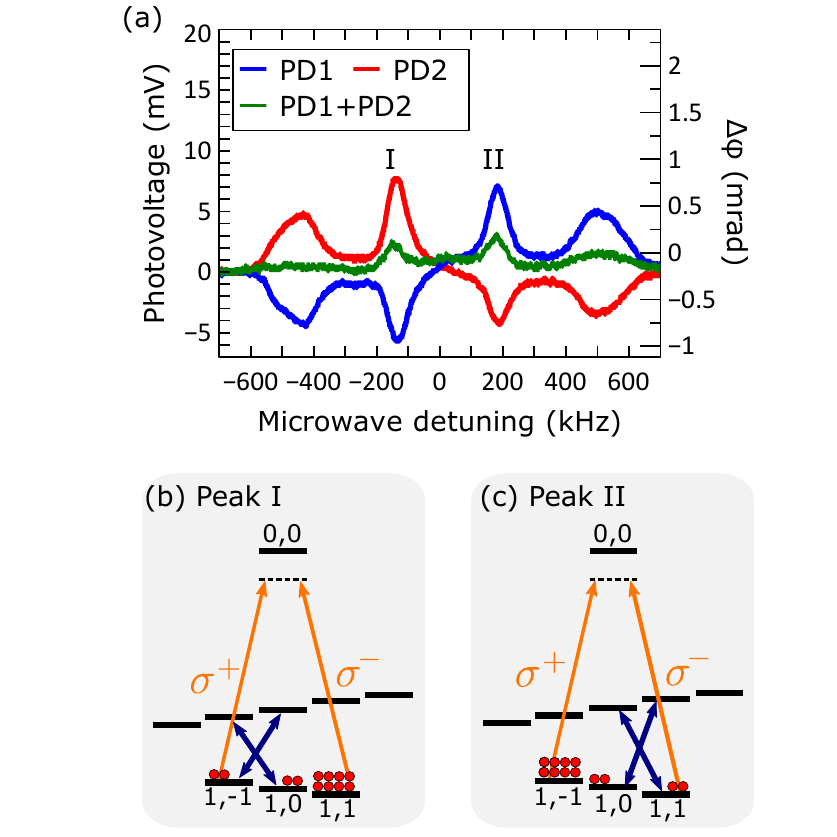}%
\caption{\label{fig:3} Dispersive measurement of MAOP. (a) Optical rotation during the microwave frequency scan detected as a change in the photovoltage on detectors PD1 and PD2. The green curve corresponds to the sum of the two voltages. Microwave detuning is given with respect to the clock transition. The microwave source is set to $-10$~dBm, the microwave frequency is scanned at a rate of $35$~Hz. The estimated optical rotation angle corresponding to peak~I is around $0.025$~mrad. (b) Schematics of the microwave (blue arrows) and probe (yellow arrow) transitions, and the expected relative distribution of the atomic population between Zeeman substates corresponding to peaks I and II from (a). The pump is not shown.}
\end{figure}

To further demonstrate the emergence of atomic polarization due to the microwave field, we explore effects in the dispersive regime. In this case, the MAOP induced atomic polarization leads to circular birefringence, resulting in the rotation of the probe polarization.
To observe the optical rotation, we use a linearly polarized probe whose frequency is set to the lower-frequency edge of the Doppler-broadened absorption peak of the D2-line originating in the $F=1$ level. In this case, the probe is red-detuned from the $\ket{F=1}\rightarrow\ket{F'=0}$ transition to the extent where its absorption is reduced and the dispersion is enhanced compared to the resonant case. In addition, $\ket{F'=1}$ and $\ket{F'=2}$ resonant classes experience an even higher detuning, and their contribution to the total signal is reduced. With a half-wave plate, we set the probe polarization at $45^\circ$  with respect to the $x$-axis ($-45^\circ$  with respect to the $z$-axis), in  which case PBS2 splits the probe power equally between the two photodetectors. In this setup, the optical rotation will result in opposite changes in the photovoltage of the two detectors without affecting the total optical power. The bias magnetic field is applied parallel to the probe, so the probe electric-field vector with respect to the quantization axis is an equal superposition of $\sigma^{+}$- and $\sigma^{-}$-polarized  fields.

The optical rotation is a result of circular birefringence due to unbalanced populations of states $\ket{1,1}$ and $\ket{1,-1}$.
Assuming that the single-atom refraction index is the same for each transition, the effective refractive index for each transition is proportional to the population at the corresponding level. As a result, the phase difference between the two polarization components leaving the ensemble is given by
\begin{align}
    \Delta \phi = kL(n_{\sigma^{+}}-n_{\sigma^{-}})=kLn[N_{-1}-N_{+1}],
    \label{eq: optical rotation angle}
\end{align}
where $k$ is the wavenumber, $L$ and the distance through which the light travels in the medium, $n$ is the single-atom refraction index, and $N_{-1}$ and $N_{+1}$ are the numbers of atoms in states $\ket{F=1,m_F=-1}$ and $\ket{F=1,m_F=1}$ states, respectively. This phase difference determines the angle by which the linear optical polarization is rotated.

Fig.~\ref{fig:3}(a) shows the relative photovoltage from PD1 and PD2  measured in the dispersive regime during a linear sweep of the microwave frequency. 
Whenever the microwave field is on resonance with an allowed hyperfine transition, it results in opposite voltage changes on the two detectors, indicating the optical rotation. The sum of the two voltages is small but not exactly zero, which indicates that some part of the signal corresponds to the absorptive regime.
The fact that the signal from each detector is antisymmetric with respect to the clock-transition frequency means that the direction of the polarization rotation depends on the sign of the microwave detuning. According to Eq.~\ref{eq: optical rotation angle}, this is a consequence of a change in the sign of the population difference between levels $\ket{1,1}$ and $\ket{1,-1}$, and thus a change in the ``direction'' of the atomic polarization. Fig.~\ref{fig:3}(b)  schematically shows the population distribution for peaks~I~and~II from Fig.~\ref{fig:3}(a). Due to a higher population of $\ket{1,1}$ in the case of negative microwave detuning, the $\sigma^{-}$-polarization component experiences a larger  refractive index and acquires a phase delay relative to the $\sigma^{+}$-polarization component, so the linear polarization rotates towards the horizontal axis, and more optical power goes to PD2. For the positive microwave detuning, the population is transferred to $\ket{1,-1}$, so  the $\sigma^{+}$ component is delayed, and the rotation direction is reversed. 
If we turn off the pump light or set its frequency off-resonance, the observed signal feature disappears, which suggests that the observed circular birefringence of the vapor results from MAOP and not from the non-linear interaction between the microwave field and the probe itself. We also highlight that when the microwave field is not on resonance with any transition, no optical rotation is observed.

Finally, we emphasize that the optical rotation due to polarization of atomic vapor discussed above corresponds to \emph{paramagnetic} Faraday rotation~\cite{Takeuchi2006}, which  differs from diamagnetic Faraday rotation, where the optical rotation angle is proportional to the magnitude of a strong static magnetic field applied parallel to the optical beam~\cite{Siddons2010}. In the case of diamagnetic Faraday rotation, transitions corresponding to different circular polarizations experience different frequency shifts due to the interaction with the static magnetic field leading to the circular birefringence.

To summarize, we introduced the technique of microwave-assisted optical pumping, which enables atomic-spin polarization engineering by combining microwave magnetic-dipole transitions with optical pumping. We demonstrated that the engineered atomic-spin polarization depends on the microwave frequency by observing its effect on optical circular birefringence and dichroism in a thermal atomic vapor in the dispersive and absorptive regimes, respectively. While the current experiments were performed in a thermal atomic vapor, and we expect better performance in cold ensembles, where collisional and spin-exchange relaxation between the ground-state levels is reduced. By engineering atomic polarization engineering via MAOP-tailored optical absorption and dispersion, we have demonstrated a suite of techniques for domain of microwave-to-optical control~\cite{Lauk2020a}, including possibilities for  a microwave-controlled polarization-selective optical switch that could encode polarization states of light, where applying a microwave field makes the atomic vapor transparent for light with a particular polarization, with applications to microwave-to-optical transduction. 

\section*{Acknowledgements}
{
This work was supported by the University of Alberta; the Natural Sciences and Engineering Research Council, Canada (Grants No. RGPIN-04523-16, No. RGPIN-2021-02884, and No. CREATE-495446-17);  the Alberta Quantum Major Innovation Fund; Alberta Innovates; and the Canada Research Chairs  (CRC) Program.
As researchers at the University of Alberta, we acknowledge that we are located on Treaty 6 territory, and that we respect the histories, languages, and cultures of First Nations, M\'etis, Inuit, and all First Peoples of Canada, whose presence continues to enrich our vibrant community.}

%


\pagebreak
\pagebreak
\newpage
\begin{widetext}

\begin{center}
\large{\bf Supplemental Materials for}\\
\large{\bf Engineering atomic polarization with microwave-assisted optical pumping}

\end{center}

\hrule




\beginsupplement

\section{Atomic populations, microwave transitions, and microwave polarizations}

To understand the principles of microwave-assisted optical pumping (MAOP), Fig.~\ref{Sup:1} illustrates distribution of atomic populations under various conditions. First, we consider the distribution of populations among the two hyperfine ground states, $\ket{F = 1}$ and $\ket{F = 2}$.  For the hyperfine splitting in $^{87}$Rb of $\omega_{\rm HFS}/2\pi  = 6.834$~GHz, the energy associated with room temperature ($T \approx 300$~K) is far greater than the energy associated with this splitting (e.g., the Boltzmann factor $\exp(-\hbar\omega_{\rm HFS}/k_{\rm B} T)\approx 1$, and we assume that the two levels are occupied according to their  degeneracy factors (associated with the $F$ quantum number), with $5/8$ of the population in $F = 2$ and $3/8$ in $F = 1$ [Fig.~\ref{Sup:1}(a)].

Next, when introducing optical pumping (as per the scheme shown in Fig.~2(b) where the optical pumping light is resonant with a transition from ground state $\ket{F = 2}$ to excited state $\ket{F' = 2}$, we assume that all population accumulate in the  lower-energy $\ket{F=1}$ hyperfine ground state and is equally distributed among Zeeman sublevels [Fig.~\ref{Sup:1}(b)]. 

Third, upon introducing a resonant microwave field, we must consider its polarization with respect to the external magnetic field $\mathbf{B}$ in a direction that sets the quantization axis for the Zeeman sublevels. If the thermal equilibration rate is negligible compared to the MAOP rate,  a resonant microwave magnetic field $\mathbf{B}_{\rm \mu}(t)$ removes the population from the optically coupled Zeeman substates, and we assume here that it splits  equally between the uncoupled states [Fig.~\ref{Sup:1}(c-e)].  This  allows us to engineer atomic polarization by deliberately choosing the microwave parameters.  For example, when the microwave magnetic field $\mathbf{B}_{\rm \mu}(t)$ is parallel to the quanitizaiton field $\mathbf{B}$, only the $\pi$-polarized microwave transitions are possible.  By tuning the microwave frequency to one of these transitions, we can empty the population from the connected ground Zeeman level [Fig.~\ref{Sup:1}(c)].  Similarly, a perpendicular microwave field effects only $\sigma^+$ and/or $\sigma^-$ transitions [Fig.~\ref{Sup:1}(d,e)], and the particular transition can be selected by frequency tuning.  Note that for the low-field linear Zeeman splitting regime in which we work, there are two sets of degenerate transitions among the nine possible transitions, yielding seven independent transition frequencies.

\clearpage
\begin{figure}[h!!!!!!]
\includegraphics{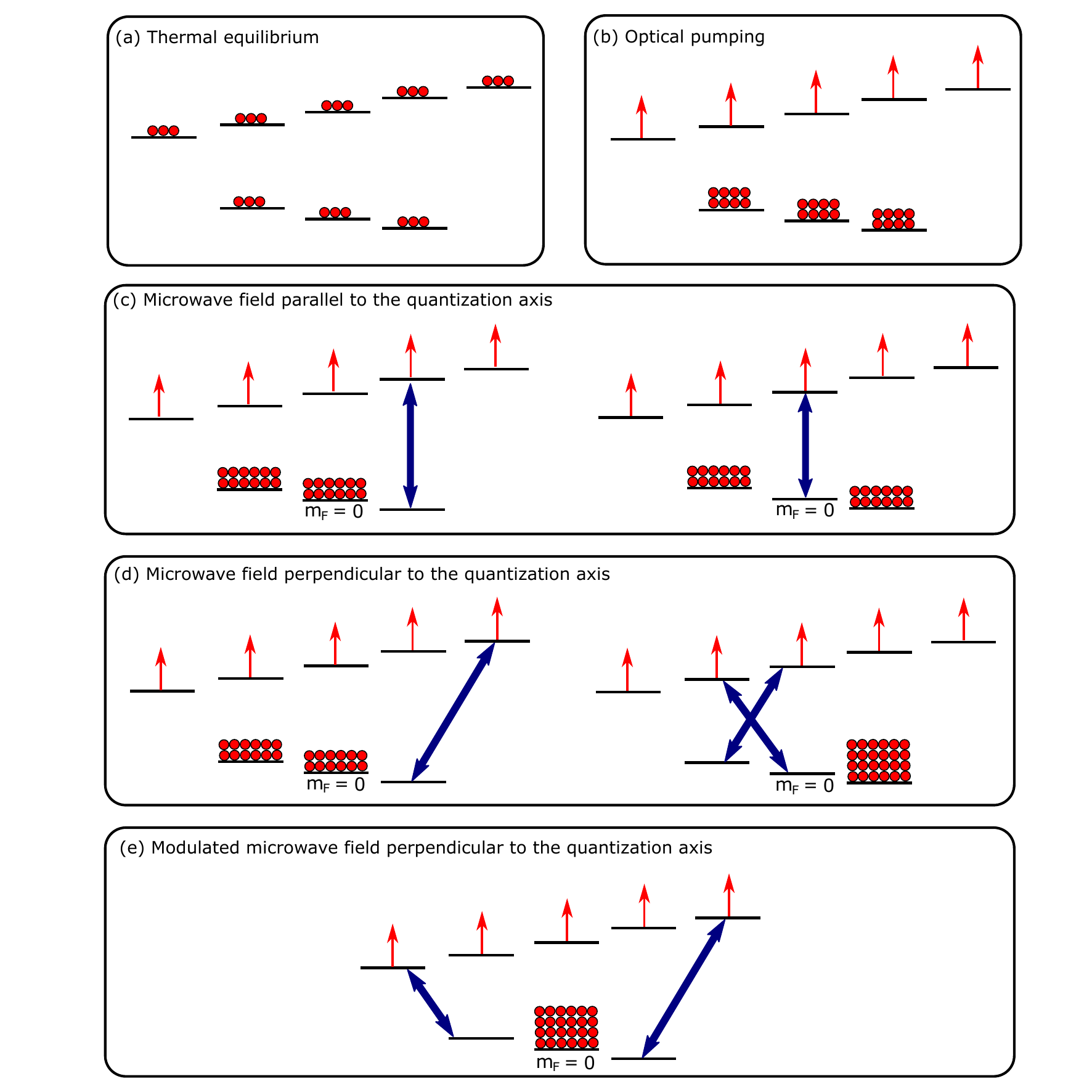}%
\caption{\label{Sup:1} Atomic population before and after the MAOP. (a) Atomic population in thermal equilibrium. (b) atomic population after optical pumping without a microwave field. (c-d) Possible atomic polarization created by the MAOP for different arrengements of an unmodulated microwave field.(e) Modulating the microwave field at twice the Larmor frequency allows to transfer the atomic population to $\ket{F=1,\mF=0}$.}
\end{figure}
\clearpage

\section{Modelling MAOP with relaxation and Doppler broadening}
The model for MAOP depicted in FIG.\ref{Sup:1}(c-e) assumes a simple configuration in which the relaxation between the ground-state levels is negligible compared to the MAOP rate, so the whole atomic population accumulates in the states uncoupled from the microwave field. Our experiments are performed in a room-temperature vapor where atomic collisions with the cell walls and thermal motion through the cell region illuminated by the pump light introduce a significant relaxation between the states. In addition, due to significant Doppler broadening, several velocity classes in the atomic ensemble are resonant with different optical transitions that contribute differently to the total absorption signal observed experimentally.  Here, we account for these effects, as well as experimental parameters like cavity linewidth, using a comprehensive model that treats optical pumping as a relaxation process to predict the relative optical absorption of a microwave-assisted optically pumped system of room-temperature atoms.

\subsection{Theoretical model}
In our model, the Hamiltonian includes eight levels corresponding to the Zeeman sublevels of the ground states of $^{87}$Rb interacting with the microwave field. 
The diagonal part of the Hamiltonian in the rotating frame (rotating at the applied microwave frequency), after applying the rotating-wave approximation, is 
\begin{align*}
   \hat{H}_0 = \hbar\sum_{\mF}\left(\dfrac{\delta}{2}-\mF\wl\right)\ketbra{F=1,\mF}{F=1,\mF}+\hbar\sum_{\widetilde{m}_{\rm F}}\left(\widetilde{m}_{\rm F}\wl-\dfrac{\delta}{2}\right)\ketbra{F=2,\widetilde{m}_{\rm F}}{F=2,\widetilde{m}_{\rm F}},
\end{align*}
where the $\wl={\mB B}/{2\hbar}$ is the Larmor frequency in a static magnetic field $\mathbf{B}$ with magnitude $B$ (already taking into account the $g_{\rm F}$ factors for $^{87}$Rb's hyperfine ground states), and $\delta$ is the detuning of the microwave field from the clock ($\ket{F = 1,\mF = 0}\rightarrow\ket{F = 2,\mF = 0}$) transition. The interaction part is given by
\begin{align*}
    \hat{V}=\dfrac{\hbar}{2}\sum_{\widetilde{m}_{\rm F}}\sum_{\mF}\Omega_{\mF,\widetilde{m}_{\rm F}}\ketbra{F=1,\mF}{F=2,\widetilde{m}_{\rm F}}+ {\rm h.c.},
\end{align*}
where $\Omega_{\mF,\widetilde{m}_{\rm F}}$ is the Rabi frequency for coupling between  between hyperfine states $\ket{F=1,\mF}$ and $\ket{F=2,\widetilde{m}_{\rm F}}$. If $\theta$ is the angle between the microwave magnetic field $\mathbf{B}_\mu$ and the quantizing field $\mathbf{B}$, the coupling strength for the $\pi$-transitions ($\mF\rightarrow\mF$) is given by 
\begin{align*}
    \Omega_{\mF,\mF}=\Omega_{0}\cos\theta\sqrt{1-g_{\rm F}^2 m_{\rm F}^2},
\end{align*}
where $\Omega_0={\mB B_\mu}/{\hbar}$, $B_\mu$ is the magnitude of the microwave magnetic field $\mathbf{B}_\mu$, and $g_{\rm F}=1/2$ is the Lande factor for \Rb. For the $\sigma^{\pm}$-transitions ($\mF\rightarrow\mF\pm1$), the coupling strength is given by
\begin{align*}
    \Omega_{\mF,\mF\pm1}=\frac{\Omega_0\sin\theta}{2I+1}\sqrt{\left(I\mp m_{\rm F}\right)^2-\dfrac{1}{4}},
\end{align*}
where $I=3/2$ is the nuclear spin of \Rb.

The finite linewidth of the cavity is taken into account in this model by modifying the coupling strength to be dependent on  detuning, and assuming this dependence has a Lorentzian character, such that
\begin{equation*}
    \Omega_{\mF,\widetilde{m}_{\rm F}}(\delta)=\left(\frac{g_{\rm C}^2}{g_{\rm C}^2+\delta^2}\right)\Omega_{\mF,\widetilde{m}_{\rm F}},
\end{equation*}
where $g_{\rm C}$ represents the cavity linewidth.

The dynamics of the system are described by the Lindblad master equation
\begin{align}
    \dfrac{d}{dt}\hat{\rho}=-\dfrac{i}{\hbar}\left[\hat{H},\hat{\rho}\right]+\dfrac{1}{2}\sum_{n} \left(2\hat{L}_n\hat{\rho} \hat{L}_n^{\dagger}-\hat{\rho}\hat{L}_n^{\dagger}\hat{L}_n-\hat{L}_n^{\dagger}\hat{L}_n\hat{\rho}\right),
    \label{eq:Lindblad}
\end{align}
where $\hat{\rho}$ is the density matrix, $\hat{H}=\hat{H}_0+\hat{V}$ is the total Hamiltonian, and $\hat{L}_{n}$ is the collapse operator due to a relaxation process $n$.

We treat optical pumping, which in practice operates through an excited state level, as a relaxation process from ground state $\ket{F=2,\widetilde{m}_{\rm F}}$ to ground state $\ket{F=1,\mF}$, and assign to it the collapse operator
\begin{align*}
    \hat{L}_{\rm OP}(\mF,\widetilde{m}_{\rm F})=\sqrt{\Gamma_{\rm OP}}\ketbra{F=1,\mF}{F=2,\widetilde{m}_{\rm F}},
\end{align*}
where $\Gamma_{\rm OP}$ is the optical pumping rate, which we assume is the same for all pairs of $\mF$ and $\widetilde{m}_{\rm F}$.

In the thermal equilibrium, we assume that all Zeeman sublevels of the ground-state hyperfine levels are equally occupied. This can be achieved by modeling the thermal relaxation with the following collapse operator:
\begin{align*}
    \hat{L}_{\rm}=\sqrt{\Gamma_{\rm th}}\ketbra{i}{j}
\end{align*}
where $\ket{i}$ and $\ket{j}$ are any two different Zeeman sublevels, and $\Gamma_{\rm th}$ is the thermal relaxation rate.

\subsection{Simulating steady-state optical absorption}
To model our system, we numerically solve Eq.~\ref{eq:Lindblad} using the computational package for Python, QuTip~[S1]. To begin with, we test the concept of the MAOP by looking at the populations $\rho_{\mF}=\bra{F=1,\mF}\hat{\rho}\ket{F=1,\mF}$  in the absence of thermal relaxation. Figs.~\ref{Sup:2} (a) and (b) show the populations corresponding to the MAOP configurations from Figs.~\ref{Sup:1}(c) and (d), respectively. These simulations show that a resonant microwave field clears out the addressed sublevels, moving the atomic population to the uncoupled sublevels. 

\begin{figure}[h!!!!!!]
\includegraphics{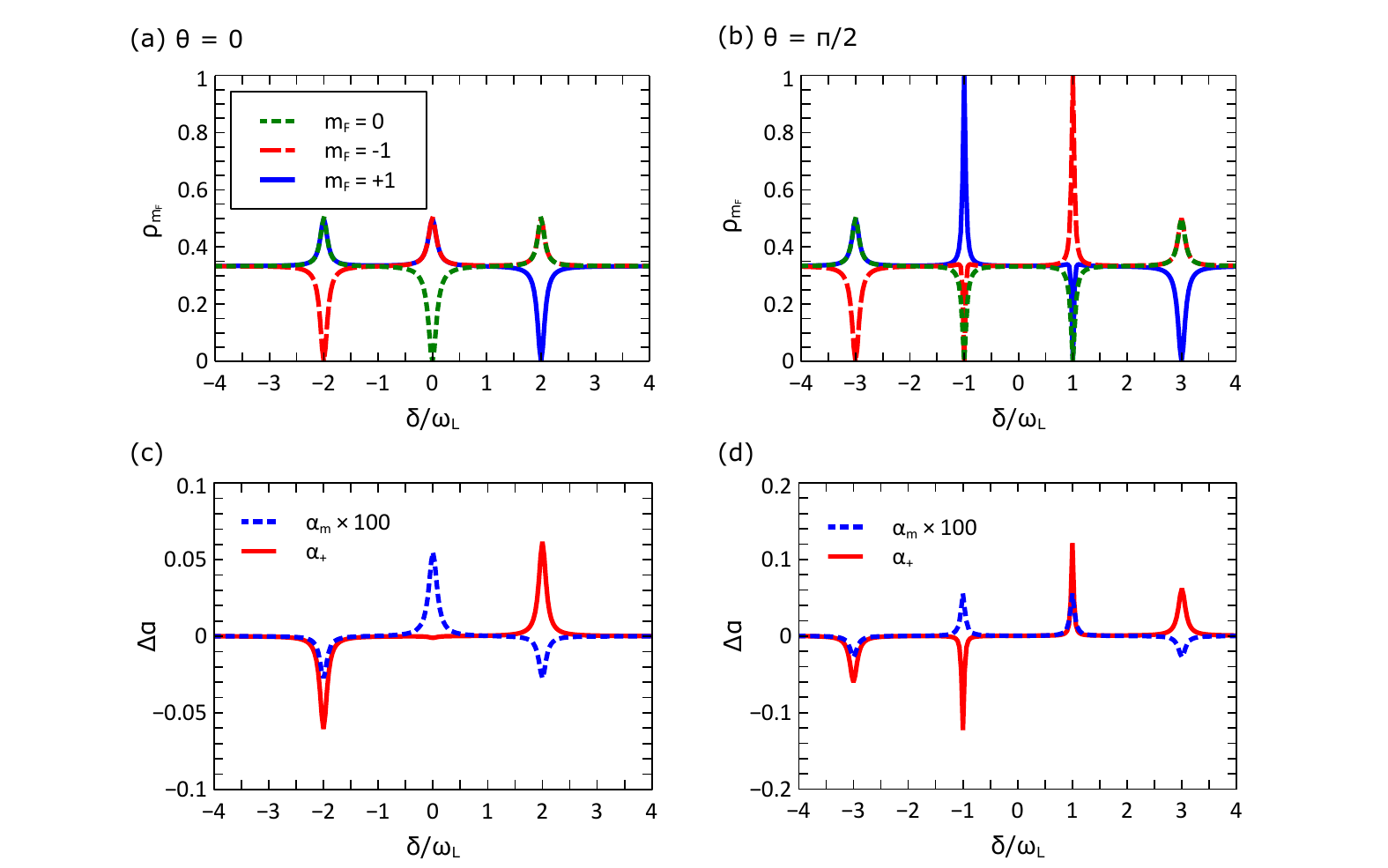}%
\caption{\label{Sup:2} Simulation of the steady-state population of $F=1$ sublevels under MAOP and the corresponding change in the absorption parameters $\alpha_{+}$ and $\alpha_{\rm m}$. The microwave detuning $\delta$ is given with the respect to the clock transition $\ket{F=1,\mF=0}\rightarrow\ket{F=2,\mF=0}$. The steady state is found using the inverse-power method~\cite[S2]. The simulation parameters are $\hbar=1$, $\Gamma_{\rm OP}=2\pi\times10^{-3}$, $\wl = 2\Gamma_{\rm OP}\times10^{5}$, $\Omega_0=\Gamma_{\rm OP}\times10^{-2}$.  (a,c) The microwave magnetic field is parallel to the quantization axis. (b,d) The microwave magnetic field is perpendicular to the quantization axis.}
\end{figure}

In order to  simulate the optical transmission shown in Fig.~3 of the main text, we first need to examine how the optical absorption depends on the distribution of the atomic population between the ground-state sublevels. For a dilute vapor, the absorption is given by
\begin{align}
    A(\omega)=1-e^{-nl\sigma(\omega)}\approx n l\sigma(\omega),
    \label{eq:Lambert law}
\end{align}
 where $n$ is the atomic density, $l$ is the length of the optical path in the medium and $\sigma(\omega)$ is the frequency-dependent absorption cross-section. For a particular $\ket{F, \mF}\rightarrow\ket{F', \mF'}$ transition, the cross-section is given by
\begin{align*}
    \sigma(\omega)=A_0 N(\omega)\| d\|^2
\end{align*}
where $A_0$ is a parameter which is the same for all transitions within the D2 line, $N(\omega)$ is the fraction of atoms undergoing this transition,
and $\| d\|$ is the dipole matrix element corresponding to this transition.
In a thermal ensemble, due to the Doppler effect, the value of $N(\omega)$ is a product of  the fraction of atoms in the velocity class resonant with the transition at frequency ($\omega$) and the  probability of an atom from this velocity class to be in the state $\ket{F,\mF}$
\begin{align*}
     N(\omega)=\sqrt{\dfrac{m}{2\pi k_B T}}\exp\left[-\left(\dfrac{c(\omega-\omega_0)}{\omega}\right)^2\dfrac{m}{2k_B T}\right]\rho_{\mF},
\end{align*}
where $m$ is the atomic mass, $T$ is the ensemble temperature, $k_{\rm B}$ is the Boltzmann constant, $c$ is the speed of light, $\omega$ is the optical laser frequency, and $\omega_0$ is the optical transition frequency in the atomic reference frame.
For the D2-line, the transition matrix element given in multiples of $\left\langle J=1 / 2 \| d\| J^{\prime}=3 / 2\right\rangle$ is
\begin{align*}
    \| d\|=\sqrt{S_{FF'\mF\mF'}}\left\langle J=1 / 2 \| d\| J^{\prime}=3 / 2\right\rangle,
\end{align*}
where $S_{FF'\mF\mF'}$ is the relative transition strength for the transition from $\ket{F,m_{\rm F}}$ to $\ket{F',m^\prime_{\rm F}}$.

As in the measurements from the main text, we consider the case for which the probe is tuned to the $\ket{F=1}\rightarrow\ket{F'=0}$ transition.  Because of Doppler broadening, three velocity classes contribute to the absorption: the zero-velocity atoms are resonant to the  $\ket{F=1}\rightarrow\ket{F'=0}$ transition itself, which we represent by the frequency $\omega_{10}$, and non-zero velocity classes are resonant to the other two transitions allowed by selection rules:  $\ket{F=1}\rightarrow\ket{F'=1}$ and $\ket{F=1}\rightarrow\ket{F'=2}$.
Because the Doppler shift for the microwave field is negligible, the MAOP equally affects the three velocity classes. However, due to different coefficients $S_{FF'\mF\mF'}$ representing the different strengths of these three transitions, each velocity class contributes to the total absorption differently. To take into account the cumulative effect that the population distribution $\rho_{\mF}$ (in the three velocity classes) has on the total absorption, we introduce a coefficient $\alpha(\omega)$, such that $A(\omega) \propto\sigma(\omega)	\propto\alpha(\omega)$, with $\alpha(\omega)$ given by
\begin{align*}
    \alpha(\omega)=\frac{1}{I}\left[I_+\alpha_+(\omega)+I_{\pi}\alpha_{\pi}(\omega)+I_-\alpha_-(\omega)\right],
\end{align*}
where $I$ is the total probe intensity and $I_{\pm}$ and $I_{\pi}$ are intensities of the $\sigma^\pm$ and $\pi$ polarization components, and 
\begin{align*}
    \alpha_{+}(\omega_{10})=\sum_{\mF}\rho_{\mF}\sum_{F'}S_{1F'\mF\mF+1}\exp\left\{-\frac{m}{2k_B T}\left[\frac{c(\omega_{10}-\omega_{1F'})}{\omega_{10}}\right]^2\right\},
\end{align*}
\begin{align*}
    \alpha_{-}(\omega_{10})=\sum_{\mF}\rho_{\mF}\sum_{F'}S_{1F'\mF\mF-1}\exp\left\{-\frac{m}{2k_B T}\left[\frac{c(\omega_{10}-\omega_{1F'})}{\omega_{10}}\right]^2\right\},
\end{align*}
\begin{align*}
    \alpha_{\pi}(\omega_{10})=\sum_{\mF}\rho_{\mF}\sum_{F'}S_{1F'\mF\mF}\exp\left\{-\frac{m}{2k_B T}\left[\frac{c(\omega_{10}-\omega_{1F'})}{\omega_{10}}\right]^2\right\},
\end{align*}
where $\omega_{FF'}$ is the frequency of the $\ket{F}\rightarrow\ket{F'}$ transition.

In our MAOP configuration, the atomic population depends on the microwave detuning $\delta$, as does, therefore, the absorption, which we now represent as $\alpha(\omega,\delta)$.
Fig.~\ref{Sup:3} shows how $\alpha(\omega_{10},\delta)$ changes due to MAOP with the microwave field detuned by $\delta$ from the clock transition. The simulation is done for the two cases of probe polarization corresponding to the  experimental data from Fig.~3 in the main text: first, a  $\sigma^{+}$-polarized probe, in which case $\alpha=\alpha_+$, and second, a mixed polarization with
\begin{align*}
    \alpha_{\rm m}(\omega,\delta)=\dfrac{1}{4}\alpha_{+}(\omega,\delta)+\dfrac{1}{4}\alpha_{-}(\omega,\delta)+\dfrac{1}{2}\alpha_{\pi}(\omega,\delta),
\end{align*}
where half of the probe power corresponds to $\pi$-polarization and the other half is split equally between $\sigma^{+}$ and $\sigma^{-}$. As seen in [Fig.~\ref{Sup:3}(a)], when there is no relaxation between the ground-state levels, MAOP should result in an antisymmetric absorption feature for a $\sigma^{+}$-polarized probe, and a symmetric feature for the probe with mixed polarization. In both cases, there is an increase in the absorption coefficient for some values of the microwave detuning, which is not what we observe experimentally, where the absorption decreases whenever the microwave field is on resonance. Upon adding thermal relaxation to the simulation results, we find the results seen in Fig.~\ref{Sup:3}(b), which is similar to what we observe experimentally.


\begin{figure}[h!!!!!!]
\includegraphics{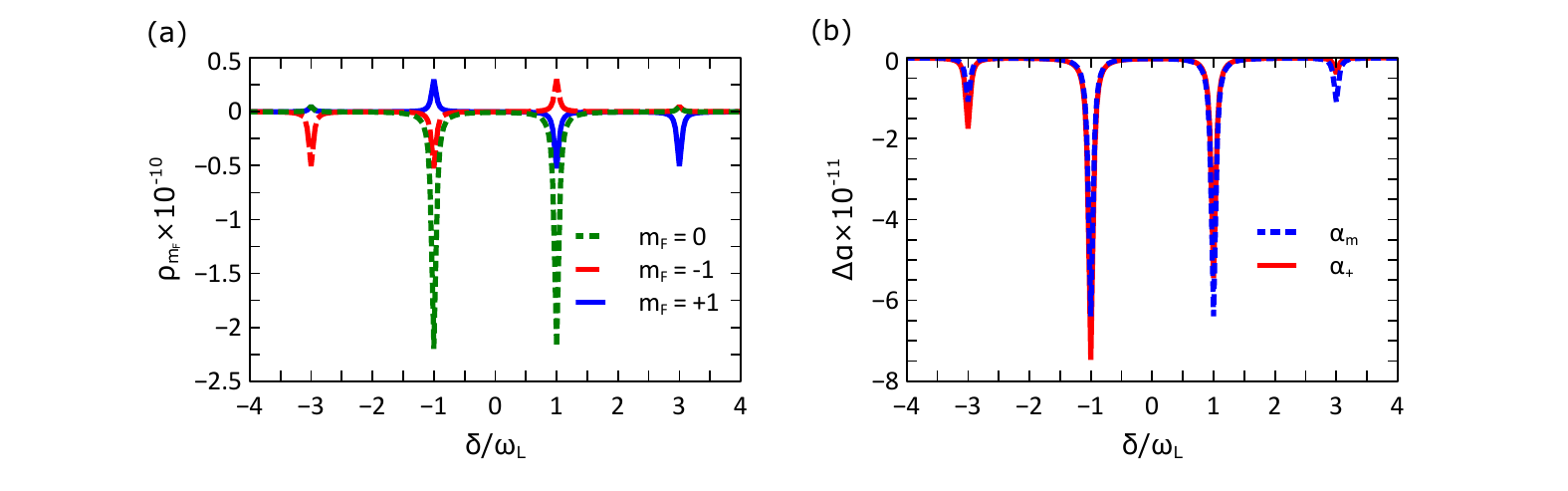}%
\caption{\label{Sup:3} Simulating the change in the steady-state absorption parameter $\alpha$ as a function of. Simulation is done with the inverse-power solution method. The simulation parameters are $\hbar=1$, $\Gamma_{\rm OP}=2\pi\times10^{-3}$, $\Omega_0=\Gamma_{\rm OP}\times10^{-2}$, $\theta=\pi/2$,  $\wl = 2\Gamma_{\rm OP}\times10^{2}$, $\Gamma_{\rm th}=\Gamma_{\rm OP}$, and $g_{C}=0.1\wl$.}
\end{figure}

\vspace{12pt}
\hrule
\vspace{12pt}
\noindent[S1] J. R. Johansson, P. D. Nation, and F. Nori, QuTiP 2:  A Python framework for the dynamics of open quantum systems, Comput.  Phys.  Commun.184, 1234 (2013).

\noindent[S2] P. D. Nation, Steady-state solution methods for open quantum optical systems, arxiv.org , 1504.06768 (2015) .
%

\begin{thebibliography}{47}
\bibitem [{SM()}]{SM}%
  \BibitemOpen
  \href@noop {} {\bibinfo {title} {See {S}upplemental {M}aterial at [url will
  be inserted by publisher] for additional information.}}\BibitemShut {Stop}%
\end{thebibliography}
%
%

\end{widetext}

\end{document}